\documentclass[nofootinbib,notitlepage,superscriptaddress,10pt,aps,prd,twocolumn]{revtex4-1}

\usepackage{amsmath,mathtools}
\usepackage{tensor}
\usepackage{amssymb}
\usepackage{graphicx}
\usepackage{verbatim}
\newcommand{\leri}[1]{\left(#1\right)}

\begin{document}

\title{$f(R)$ gravity with Torsion and the Immirzi field:
signature for GW detection}
\author{Flavio Bombacigno}
\email{flavio.bombacigno@uniroma1.it}
\affiliation{Physics Department, ``Sapienza'' University of Rome, P.le Aldo Moro 5, 00185 (Roma), Italy}
\author{Giovanni Montani}
\email{giovanni.montani@enea.it}
\affiliation{ENEA, FSN-FUSPHY-TSM, R.C. Frascati, Via E. Fermi 45, 00044 Frascati, Italy.\\
Physics Department, ``Sapienza'' University of Rome, P.le Aldo Moro 5, 00185 (Roma), Italy}

\begin{abstract}
We propose a Nieh-Yan extension of $f(R)$ models in the presence of Immirzi field and the resulting theory is studied in the context of first order formalism. We claim as in our model the ambiguities related to the Palatini approach are solved, allowing us to fully solve the theory in terms of the Immirzi field. In particular, we analyze within such a theoretical framework the propagation  of gravitational waves in vacuum and outline as a dynamical Immirzi field could be regarded as a propagating torsion field with non-trivial and detectable effects on standard gravitational polarizations.
\end{abstract}

\maketitle

\section{Introduction}
General Relativity is a well-established 
theory with respect its "kinematical'' formulation, i.e. the geometrical
nature of the gravitational interaction, together with the implementation of the tensorial language (General Relativity Principle) \cite{Gravitation}. However, the dynamical setting of Einsteinian gravity is not equivalently settled down, in fact the Einstein-Hilbert action \cite{CanonicalQG} just corresponds to the simplest choice. 
\\ The main important feature of the Einstein equations is that they contain at most second order derivatives in the metric tensor, in which they are also linear, but it must be recalled that Einsteinian gravity has some important experimental validations \cite{Ni:2016dwy}. 
\\ \indent Among the possible extensions of General Relativity, the so-called $f(R)$ theory has been widely studied \cite{Sotiriou:2008rp,Nojiri:2010wj,Nojiri:2017ncd} for 
its simplicity (only a scalar degree of freedom is added) and generality 
(a set of fourth order Einstein equations can be easily written down). 
\\ The motivation for such extended formulations, which must pass the Solar System test \cite{Zakharov:2006uq}, must be individualized not only in theoretical requirements, but also in the perspective of explaining 
till unidentified astrophysical and cosmological effects, like dark matter and dark energy \cite{Primordial}. 
\\ \indent Another important ambiguity of the gravitational field formulation consists in the possibility to link 
the affine connection to the metric variables before the variational principle is performed (second order formulation \cite{Landau}), 
or in taking the affine connection as an independent field (first order or
Palatini formulation \cite{Palatini}). 
\\ When the Einstein-Hilbert action is considered in vacuum, it is easy to demonstrate the equivalence of the second order and the Palatini formulation: the dependence of the affine connection is 
recovered from the field equations and the Einstein picture of gravitation is then naturally reconstructed \cite{Wald}. 
\\ \indent However, when matter is included in the 
dynamical problem the situation significantly changes: using spin-connection variables \cite{Cartan1, Cartan2}, it is rather simple to 
verify that the second order formulation of gravitation in the presence of
spinor fields is no longer equivalent to the Palatini method. In fact, the first Cartan equation, obtained by varying the total action with respect the spin connections, predicts the emergence of non-zero torsion (antisymmetric affine connection component \cite{Hehl:1976kj, Shapiro:2001rz}), absent, by
construction in the second order metric approach \cite{CanonicalQG}. 
\\ \indent The situation becomes even more intricate when the $f(R)$ theory of gravitation is formulated in the Palatini scheme. In fact, as shown in \cite{Sotiriou:2008rp}, the vacuum theory is just allowed for the $R^2$ 
gravity and the general case is available in the presence of matter only, having care to impose the non-trivial constraint that the matter action does not couple with the torsion field. 
\\ \indent In the present analysis we consider such a very subtle question, by addressing the following point of view: 
if the $f(R)$ model appears to be consistent only removing any coupling of matter with the torsion field (this is necessary for a suitable redefinition of the gravitational variable), 
we infer that this fact is simply marking the necessity to include torsion 
in the theory since from the very beginning. In particular, the torsion field must appear in the gravitation-matter action with its known (not simply $f(R)$, with non-Riemannian $R$) contributions. 
\\ \indent Our proposed model couples to the gravitation-matter action, associated to a generic $f(R)$ scheme, a topological term (the so-called Nieh-Yan term \cite{NY1, NY2}) time a scalar field, ensuring a real contribution. 
\\ By other words, in the language of the Loop Quantum Gravity (LQG) formulation \cite{Thiemann:2007zz}, we are adding to the $f(R)$ model, 
a generalization of the Holst contribution \cite{Holst:1995pc}, but associated to an Immirzi field, instead of a parameter \cite{Immirzi:1996di, Immirzi:1996dr, Rovelli:1997na, TorresGomez:2008fj, Calcagni:2009xz, Cianfrani:2009sz, Bombacigno:2016siz}. 
\\ \indent We demonstrate how this reformulation of 
the Palatini method for the $f(R)$ theories is now at all viable in vacuum for a generic form of $f(R)$. Furthermore, we can develop a consistent theory even when matter action is directly coupled to the torsion field. 
\\ In practice, we are able to eliminate the torsion field from the dynamics, by re-substituting its expression, in terms of the metric variable and the Immirzi field, into the equations of motion (or action), so obtaining a new formulation, which is, in principle, easier to be addressed. 
\\ The achievement of this new viable formulation confirms the validity of our guess about the necessity to include the torsion field \textit{ab initio} in the variational principle. Furthermore, we analyze in some detail the behaviour of the gravitational waves in the obtained framework, 
and one of the main merit of the present analysis consists of the emergence of a clear phenomenological mark of the gravitational wave polarization in the considered model.
\\ \noindent Indeed, we demonstrate that the main axes of the cross-like polarization are no longer orthogonal, as in General Relativity, but they form an angle depending on the details of the addressed theory: this feature can be, in principle, identified in the data analyses from the interferometers LIGO and VIRGO \cite{Abbott:2016blz, Abbott:2016nmj, Abbott:2017vtc, Abbott:2018utx, Abbott:2017tlp}. 
\\ \indent The paper is organized as follows. In Sec. \ref{sec2} we briefly review $f(R)$ theories within the first order formalism, outlining its critical issues. In Sec. \ref{sec3} we present our model, discussing the strategy for solving the ambiguities stemming in Sec. \ref{sec2}. In Sec. \ref{sec4} we analyze the effects of torsion due to the Immirzi field on the propagation of gravitational waves. Eventually, in Sec. \ref{sec5} conclusions follow.
\section{Palatini approach to $f(R)$ theories}\label{sec2}
When one adopts a first order formalism, that is the Palatini scheme, the affine connection $\Gamma^{\rho}_{\;\;\mu\nu}$ and the metric $g_{\mu\nu}$ are considered independent variables. Thus, the Riemann tensor $R_{\mu\nu\rho\sigma}$ is formally defined like a function of the connection only, \textit{i.e.}:
\begin{equation}
R^{\mu}_{\;\;\nu\rho\sigma}=\partial_{\rho}\Gamma^{\mu}_{\;\;\nu\sigma}-\partial_{\sigma}\Gamma^{\mu}_{\;\;\nu\rho}+\Gamma^{\mu}_{\;\;\tau\rho}\Gamma^{\tau}_{\;\;\nu\rho}-\Gamma^{\mu}_{\;\;\tau\sigma}\Gamma^{\tau}_{\;\;\nu\rho},
\end{equation}
whence the expression for the Ricci scalar 
\begin{equation}
R=g^{\mu\nu}R_{\mu\nu}(\Gamma)=g^{\mu\nu}R\indices{^{\rho}_{\mu\rho\nu}}(\Gamma).
\label{RicciP}
\end{equation}
Therefore, the action (in units where $\hbar=c=8\pi G=1$) for a generic $f(R)$ theory in the presence of matter reads:
\begin{equation}
S_{P}=\frac{1}{2}\int d^4x\;\sqrt{-g}f(R)+S_M(g_{\mu\nu},\psi),
\label{actionP}
\end{equation}
where $f(R)$ is a scalar function of the Ricci curvature $R$ and $S_M$ denotes the action for the matter fields $\psi$, which we assume not depending on connection $\Gamma\indices{^\rho_{\mu\nu}}$.
\\ \indent Varying \eqref{actionP} with respect to the metric and the connection carries out, respectively:
\begin{equation}
f'(R)R_{(\mu\nu)}-\frac{1}{2}f(R)=T_{\mu\nu}
\label{gP}
\end{equation}
and
\begin{equation}
\nabla_{\sigma}\leri{\sqrt{-g}f'(R)g^{\mu\nu}}-\nabla_{\rho}\leri{\sqrt{-g}f'(R)g^{\rho(\nu}}\delta^{\mu)}_{\sigma}=0,
\label{connectP}
\end{equation}
where round brackets denote symmetrization on the indices, $\nabla_{\mu}$ is the covariant derivative from $\Gamma\indices{^\rho_{\mu\nu}}$ \textbf{and} the stress-energy tensor $T_{\mu\nu}$ is given by
\begin{equation}
T_{\mu\nu}\equiv -\frac{2}{\sqrt{-g}}\frac{\delta S_M}{\delta g^{\mu\nu}}.
\end{equation}
Tracing \eqref{connectP} yields
\begin{equation}
\nabla_{\mu}\leri{\sqrt{-g}f'(R)g^{\mu\nu}}=0,
\end{equation}
that reinserted in \eqref{connectP} itself leads to the equivalent equation
\begin{equation}
\nabla_{\rho}\leri{\sqrt{-g}f'(R)g^{\mu\nu}}=0.
\label{connectP2}
\end{equation}
Condition \eqref{connectP2} can be formally restated as the Levi-Civita definition for the connection $\Gamma\indices{^\rho_{\mu\nu}}$, provided we conformally rescale the metric $g_{\mu\nu}$ by means of
\begin{equation}
\tilde{g}_{\mu\nu}\equiv f'(R)g_{\mu\nu},
\end{equation}
whence
\begin{equation}
\nabla_{\rho}\leri{\sqrt{-\tilde{g}}\;\tilde{g}^{\mu\nu}}=0,
\end{equation}
which admits the well-known solution
\begin{equation}
\Gamma\indices{^\rho_{\mu\nu}}=\frac{1}{2}\tilde{g}^{\rho\sigma}\leri{\partial_{\nu}\tilde{g}_{\mu\sigma}
+\partial_{\mu}\tilde{g}_{\nu\sigma}-\partial_{\rho}\tilde{g}_{\mu\nu}}.
\end{equation}
It is worth noting that such a procedure is meaningful only after having express $R$ as a function of quantities $\Gamma$-independent. This can be achieved by taking the trace of the equation for the gravitational field, which turns out to be just an algebraic relation, once the $f(R)$ model has been fixed, linking the Ricci scalar $R$ to the trace of stress-energy tensor $T$, \textit{i.e.}:
\begin{equation}
f'(R)R-2f(R)=T.
\label{T1}
\end{equation}
However, we outline as in the vacuum case, that is for vanishing $T$, such an equation admits as unique solution the case $f(R)=a R^2$, where $a$ is a constant. Moreover, we note that the entire machinery relies on the assumption that matter does not depend on the connections, consisting also in torsion, ruling out in such a way the presence of spinors, which instead naturally couple to connections. In general, therefore, the procedure can be applied to very few $f(R)$ models and the theory can not be completely solved.

\section{Nieh-Yan extension of $f(R)$ theories}\label{sec3}
The starting point of our analysis is an $f(R)$ extension of the Nieh-Yan action \cite{NY1, NY2} and for the sake of simplicity we shall consider the vacuum case, that is:
\begin{equation}
\begin{split}
S&=\frac{1}{2}\int d^4x\;\sqrt{-g}f(R)\;+\\
&+\frac{1}{4}\int d^4x\;\sqrt{-g}\left(\beta(x)\epsilon^{\mu\nu\rho\sigma}\left(g_{\tau\lambda}T\indices{^{\tau}_{\mu\nu}}T\indices{^{\lambda}_{\rho\sigma}}-R_{\mu\nu\rho\sigma}\right)\right),
\label{NYaction}
\end{split}
\end{equation}
being $\beta(x^{\mu})$ the reciprocal of the Immirzi field and $T^{\tau}_{\;\;\mu\nu}$ the torsion tensor, given by:
\begin{equation}
T^{\rho}_{\;\;\mu\nu}=\Gamma^{\rho}_{\;\;\mu\nu}-\Gamma^{\rho}_{\;\;\nu\mu},
\end{equation}
where $\Gamma\indices{^\rho_{\mu\nu}}$ is an independent variable with respect to metric, as before.
In particular, $\Gamma^{\rho}_{\;\;\mu\nu}$ can be put in the form
\begin{equation}
\Gamma^{\rho}_{\;\;\mu\nu}=\bar{\Gamma}^{\rho}_{\;\;\mu\nu}+K^{\rho}_{\;\;\mu\nu},
\label{GK}
\end{equation}
with $\bar{\Gamma}^{\rho}_{\;\;\mu\nu}$ denoting the Levi-Civita torsionless connection\footnote{In the following we denote with a bar the torsionless quantities.}, depending on metric variable and its derivatives only, and $K^{\rho}_{\;\;\mu\nu}$ the contorsion tensor, related to the torsion tensor by
\begin{equation}
K^{\rho}_{\;\;\mu\nu}=\frac{1}{2}\left(T\indices{^{\rho}_{\mu\nu}}-T\indices{_{\mu}^{\rho}_{\nu}}-T\indices{_{\nu}^{\rho}_{\mu}}\right).
\label{contorsion}
\end{equation}
Accordingly to \cite{Calcagni:2009xz}, we decompose the torsion tensor into its irreducible representations
\begin{equation}
T_{\mu\nu\rho}=\frac{1}{3}\left(T_{\nu}g_{\mu\rho}-T_{\rho}g_{\mu\nu}\right)-\frac{1}{6}\epsilon_{\mu\nu\rho\sigma}S^{\sigma}+q_{\mu\nu\rho}
\label{torsion}
\end{equation}
with $T_{\mu}=T^{\nu}_{\;\;\mu\nu}$ the trace vector, $S_{\sigma}=\epsilon_{\mu\nu\rho\sigma}T^{\mu\nu\rho}$ the pseudotrace axial vector and $q_{\mu\nu\rho}$ a completely antisymmetric traceless tensor, that is \begin{equation}
\begin{split}
q^{\mu}_{\;\;\nu\mu}&=0 \\ 
\epsilon_{\mu\nu\rho\sigma}q^{\mu\nu\rho}&=0.
\end{split}
\end{equation}
\\ \indent Inserting \eqref{contorsion} and \eqref{torsion} in \eqref{NYaction} leads us to the equivalent action
\begin{equation}
\mathcal{S}=\;\frac{1}{2}\int d^4x\sqrt{-g}\;f(R)\;+\frac{1}{4}\int d^4x\sqrt{-g}\;\beta(x)\bar{\nabla}_{\mu}S^{\mu},
\label{NYequiv}
\end{equation}
where $R$ has to be now understood as a function of $\bar{R},\;T_{\mu},\;S_{\mu},\;q_{\mu\nu\rho}$, namely:
\begin{equation}
R= \bar{R}+\frac{1}{24}S_{\mu}S^{\mu}-\frac{2}{3}T_{\mu}T^{\mu}+\frac{1}{2}q_{\mu\nu\rho}q^{\mu\nu\rho}-2\bar{\nabla}_{\mu}T^{\mu}
\label{R}
\end{equation}
and $\bar{\nabla}$ denotes the torsionless covariant derivative with respect to Levi-Civita connection.
It is worth noting that \eqref{NYequiv} is still a truly first order formulation of the theory. Indeed, the terms in \eqref{NYequiv} depending on $\left\{T_{\mu},\;S_{\mu},\;q_{\mu\nu\rho}\right\}$ can be considered as the remnant of the independent part of the connection, once the splitting \eqref{GK} is performed.
\\ Now, varying \eqref{NYequiv} with respect to the metric field $g_{\mu\nu}$ yields
\begin{equation}
\begin{split}
&f'(R)\bar{R}_{\mu\nu}-\bar{\nabla}_{\mu}\bar{\nabla}_{\nu}f'(R)-g_{\mu\nu}\left(\frac{1}{2}f(R)-\bar{\Box}f'(R)\right)+\\
&+f'(R)\left(\frac{1}{24}S_{\mu}S_{\nu}-\frac{2}{3}T_{\mu}T_{\nu}+\frac{3}{2}q_{\mu\rho\sigma}q\indices{_{\nu}^{\rho\sigma}}\right)+\\
&-\frac{1}{2}S_{\mu}\bar{\nabla}_{\nu}\beta(x)+\frac{1}{4}g_{\mu\nu}(S_{\rho}\bar{\nabla}^{\rho}\beta(x))+\\
&+2T_{\mu}\bar{\nabla}_{\nu}f'(R)-g_{\mu\nu}\bar{\nabla}_{\rho}(f'(R)T^{\rho})=0,
\label{eqG}
\end{split}
\end{equation}
with a prime denoting differentiation with respect to the argument, whereas the equations for the torsion components $\left\{T_{\mu},\;S_{\mu},\;q_{\mu\nu\rho}\right\}$ and the scalar field $\beta(x)$, turn out to be, respectively:
\begin{equation}
T_{\mu}=\frac{3}{2}\frac{\bar{\nabla}_{\mu}f'(R)}{f'(R)}\quad
S_{\mu}=6\frac{\bar{\nabla}_{\mu}\beta(x)}{f'(R)}\quad
q_{\mu\nu\rho}=0
\label{TT}
\end{equation}
and
\begin{equation}
\bar{\nabla}_{\mu}S^{\mu}=0.
\label{B}
\end{equation}
We stress the fact that the system given by \eqref{eqG},\eqref{TT},\eqref{B} is not completely determined. In fact, the presence of terms proportional to $f'(R)$ in the equations of motion for the components of torsion prevent us to fully solve it, being the relation between $R$ and $\left\{T_{\mu},\;S_{\mu},\;q_{\mu\nu\rho}\right\}$ self-recursive, as it can be seen from \eqref{R}.
\\ Likewise we did in the previous section for the simple $f(R)$ model, it is possible to express $R$ in terms of quantities which do not depend on the connection. Indeed, once we have substituted the expressions for $T_{\mu},\;S_{\mu},\;q_{\mu\nu\rho}$ in \eqref{eqG} and after a bit of algebraic manipulation, the relation \eqref{T1} is replaced by a quite similar expression containing the kinetical term for the Immirzi field instead of the trace of the stress-energy tensor, that is:
\begin{equation}
f'(R)^2R-2f(R)f'(R)=-3\bar{\nabla}_{\mu}\beta(x)\bar{\nabla}^{\mu}\beta(x).
\label{T2}
\end{equation}
Although in \eqref{NYaction} $\beta(x^{\mu})$ was coupled to the torsion, namely connection, we outline how such a relation does not suffer from ambiguities, being well-defined also in the vacuum case. Moreover, when $\beta(x)$ is kept constant, as in the standard formulation of LQG, the \eqref{T1} is recovered, provided $f'(R)\neq 0$.
\\ \indent Concerning the equations for $g_{\mu\nu}(x)$ and $\beta(x)$, they can be rearranged like
\begin{equation}
\begin{split}
\bar{R}_{\mu\nu}&=g_{\mu\nu}\frac{f(R)+\bar{\Box}f'(R)}{2f'(R)}+\frac{\bar{\nabla}_{\mu}\bar{\nabla}_{\nu}f'(R)}{f'(R)}+\\
&-\frac{3}{2}\frac{\bar{\nabla}_{\mu}f'(R)\bar{\nabla}_{\nu}f'(R)}{f'(R)^2}+\\
&+\frac{3}{2}\frac{\bar{\nabla}_{\mu}\beta(x)\bar{\nabla}_{\nu}\beta(x)}{f'(R)^2}-g_{\mu\nu}\frac{3}{2}\frac{\bar{\nabla}_{\rho}\beta(x)\bar{\nabla}^{\rho}\beta(x)}{f'(R)^2}
\end{split}
\label{gf}
\end{equation}
and
\begin{equation}
\bar{\Box}\beta(x)=\bar{\nabla}_{\mu}\beta(x)\bar{\nabla}^{\mu}\ln f'(R).
\label{beta}
\end{equation}
We conclude this section noting that the same results could be obtained even if we substituted the relations \eqref{TT} at lagrangean level. Furthermore, to solve simultaneously \eqref{gf} and \eqref{beta}, assuming that we have solve \eqref{T2} for $R$, is in general an highly non-trivial task also for simple $f(R)$ models. It motivated us to seek for simpler scenarios, as those the linearized theory can offer, with the aim of investigating new effects in gravitational waves propagation due to the Immirzi field.

\section{Gravitational waves in the presence of the Immirzi field}\label{sec4}
\begin{figure}
\begin{center}
\includegraphics[scale=0.6]{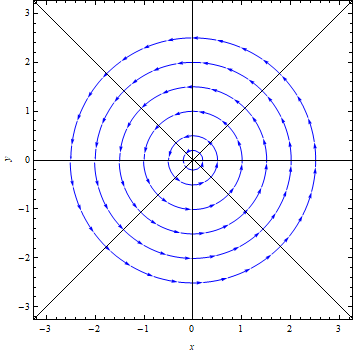}
\caption{The lines of the tangential forces related to the $\beta$-mode for $\sin\omega t >0$ (the direction of arrows reverse for $\sin\omega t<0$). With the black solid lines are shown the standard polarizations ($\epsilon=0$)}
\label{graficobeta}
\end{center}
\end{figure}

\begin{figure}
\begin{center}
\includegraphics[scale=0.6]{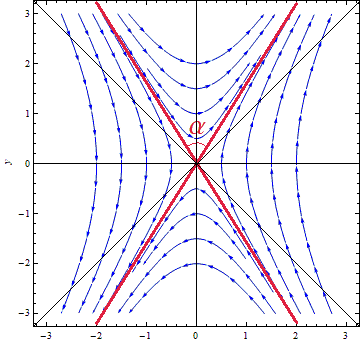}
\caption{The lines of force related to the case $0<\epsilon<1$ for $\sin\omega t >0$ (the direction of arrows reverse for $\sin\omega t<0$). With the black solid lines are shown the standard polarizations ($\epsilon=0$), whereas with the red solid lines the effective cross mode.}
\label{graficoeless1}
\end{center}
\end{figure}

\begin{figure}
\begin{center}
\includegraphics[scale=0.6]{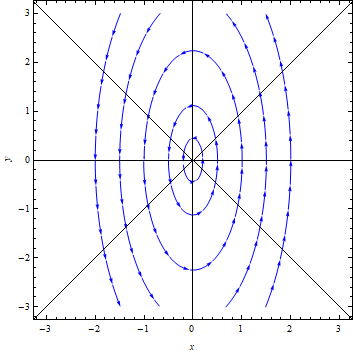}
\caption{The lines of force related to the case $\epsilon>1$ for $\sin\omega t >0$ ( the direction of arrows reverse for $\sin\omega t<0$). With the black solid lines the standard polarizations are shown.}
\label{graficoemore1}
\end{center}
\end{figure}

\begin{figure}
\begin{center}
\includegraphics[scale=0.6]{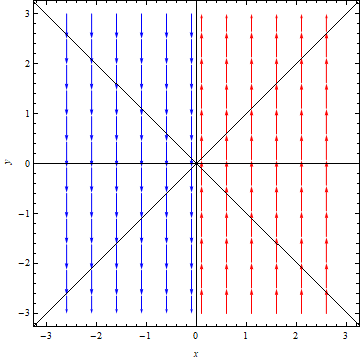}
\caption{The lines of force related to the case $\epsilon=1$ for $\sin\omega t >0$ (the direction of arrows reverse for $\sin\omega t<0$). With the black solid lines the standard polarizations are shown, whereas the coloured solid lines represent the strain stresses.}
\label{graficoe1}
\end{center}
\end{figure}

In order to study the polarization modes of gravitational waves within a general $f(R)$ scheme, it is useful to recast the theory in the form of scalar-tensor model \cite{ST}. In particular, it can be shown that \eqref{NYaction} is equivalent to the effective action
\begin{equation}
S_J=\frac{1}{2}\int d^4x\;\sqrt{-g}\leri{\phi \bar{R}-g^{\mu\nu}K_{\mu\nu}(\phi,\beta)-V(\phi)},
\label{Scalartensor}
\end{equation}
where $\phi\equiv f'(R)$ and
\begin{equation}
\begin{split}
&K_{\mu\nu}(\phi,\beta)=\frac{3}{2\phi}\leri{\bar{\nabla}_{\mu}\beta\bar{\nabla}_{\nu}\beta-\bar{\nabla}_{\mu}\phi\bar{\nabla}_{\nu}\phi}\\
&V(\phi)\equiv \phi R(\phi)-f(R(\phi)).
\end{split}
\end{equation}
Varying \eqref{Scalartensor} with respect to $g_{\mu\nu}$ carries out:
\begin{equation}
\begin{split}
\bar{R}_{\mu\nu}-\frac{1}{2}g_{\mu\nu}\bar{R}=&+\frac{1}{\phi}K_{\mu\nu}(\phi,\beta)\;+\\&-\frac{1}{2\phi}g_{\mu\nu}\leri{K_{\;\;\rho}^{\rho}(\phi,\beta)+V(\phi)}+\\
&+\frac{1}{\phi}\leri{\bar{\nabla}_{\mu}\bar{\nabla}_{\nu}\phi-g_{\mu\nu}\bar{\Box}\phi},
\label{scalartensorg}
\end{split}
\end{equation}
while the equations for $\phi$ and $\beta$ are given by, respectively:
\begin{equation}
\bar{R}=-\frac{3}{2\phi^2}\leri{\bar{\nabla}_{\mu}\beta\bar{\nabla}^{\mu}\beta+\bar{\nabla}_{\mu}\phi\bar{\nabla}^{\mu}\phi}+\frac{3\bar{\Box}\phi}{\phi}+V'(\phi)
\label{scalartensorphi}
\end{equation}
and
\begin{equation}
\bar{\Box}\beta(x)=\frac{\bar{\nabla}_{\mu}\beta(x)\bar{\nabla}^{\mu}\phi}{\phi}.
\end{equation}
Now, if we consider metric perturbations around Minkowski background
\begin{equation}
g_{\mu\nu}=\eta_{\mu\nu}+h_{\mu\nu},
\end{equation}
to the first order in $h_{\mu\nu}$ the torsionless Riemann tensor and the Ricci tensor read as, respectively:
\begin{equation}
\begin{split}
\bar{R}_{\rho\sigma\mu\nu}&=\frac{1}{2}\leri{\partial_{\sigma}\partial_{\mu}h_{\rho\nu}+\partial_{\rho}\partial_{\nu}h_{\sigma\mu}-\partial_{\sigma}\partial_{\nu}h_{\rho\mu}-\partial_{\rho}\partial_{\mu}h_{\sigma\nu}}\\
\bar{R}_{\mu\nu}&=\frac{1}{2}\leri{\partial_{\mu}\partial_{\rho}h\indices{^{\rho}_{\nu}}+\partial_{\nu}\partial_{\rho}h\indices{^{\rho}_{\mu}}-\partial_{\mu}\partial_{\nu}h-\bar{\Box}h_{\mu\nu}},
\end{split}
\label{Riccit1or}
\end{equation}
where $h\equiv \eta^{\mu\nu}h_{\mu\nu}$ is the trace, whence the Ricci scalar
\begin{equation}
\bar{R}=\eta^{\mu\nu}\bar{R}_{\mu\nu}=\partial_{\mu}\partial_{\nu}h^{\mu\nu}-\bar{\Box}h.
\label{Riccis1or}
\end{equation}
Regarding the scalar fields $\phi$ and $\beta$, they can be rewritten as: 
\begin{equation}
\begin{split}
&\phi=\phi_0+\delta\phi\\
&\beta=\beta_0+\delta\beta,
\end{split}
\label{scalarpert}
\end{equation}
where $\delta\beta,\,\delta\phi$ are order $O(h)$, $\beta_0$ a constant and $\phi_0$ a steady minimum for the potential $V(\phi)$ \cite{Corda}, that is:
\begin{equation}
\begin{split}
&V(\phi)\simeq \frac{1}{2}m\delta\phi^2\\
&V'(\phi)\simeq m\delta\phi,
\end{split}
\end{equation}
being $m$ a constant. Then, taking in account \eqref{scalarpert}, the equation for the gravitational field \eqref{scalartensorg} up to first order reduces to
\begin{equation}
R_{\mu\nu}-\frac{1}{2}\eta_{\mu\nu}R=\partial_{\mu}\partial_{\nu}\tilde{\phi}
-\eta_{\mu\nu}\bar{\Box}\tilde{\phi},
\label{GWg}
\end{equation}
with $\tilde{\phi}\equiv\frac{\delta\phi}{\phi_0}$. Eventually, considering the expansions \eqref{Riccit1or}, \eqref{Riccis1or} and generalizing \cite{Berry:2011pb,Rizwana:2016qdq} the trace-reverse tensor $\tilde{h}_{\mu\nu}$ by means of
\begin{equation}
h_{\mu\nu}\equiv\tilde{h}_{\mu\nu}-\eta_{\mu\nu}\leri{\frac{\tilde{h}}{2}-\tilde{\phi}},
\label{tracciainvFR}
\end{equation}
where $\tilde{h}=\eta^{\mu\nu}\tilde{h}_{\mu\nu}$, it can be shown, once the Lorenz gauge $\partial^{\mu}\tilde{h}_{\mu\nu}=0$ is fixed, that \eqref{GWg} can be put in the form:
\begin{equation}
\bar{\Box}\tilde{h}_{\mu\nu}=0.
\label{GWonda}
\end{equation}
As it is outlined in \cite{Rizwana:2016qdq}, within the $f(R)$ framework the traceless condition can not be simultaneously imposed with the Lorenz gauge, giving rise to an additional massless scalar mode in $\tilde{h}_{\mu\nu}$, called breathing mode, besides the well known tensorial polarizations $+,\times$.
\\ \indent Moreover, combining the equation for $\delta\phi$ with the trace of the gravitational field equation generally highlights a further massive mode and we usually expect four polarizations. However, in our case if we follow such a procedure, \textit{i.e.} we compare the trace of \eqref{scalartensorg} with \eqref{scalartensorphi}, it turns out that the scalar perturbation $\delta\phi$ has to vanish identically.
Therefore, as it can be seen by \eqref{GWg} and \eqref{tracciainvFR}, this condition compels the breathing mode to disappear as well and allows us to rescue the traceless condition on $\tilde{h}_{\mu\nu}$. Thus, the two standard polarizations are restored, in agreement with the Palatini formulation of gravitational waves \cite{Ferraris:1992dx,Naf:2008sf}.
\\ Furthermore, since the geometrical scalar perturbation $\delta\phi$ is forced to vanish in linearized theory, the contorsion tensor simply reduces to a completely antisymmetric tensor, that is:
\begin{equation}
K_{\mu\nu\rho}=-\frac{1}{2\phi_0}\epsilon_{\mu\nu\rho\sigma}\partial^{\sigma}\delta\beta.
\label{Kpert}
\end{equation}
Then, given the form \eqref{Kpert}, it is easy to see how the motion of a free falling particle, thought as the auto-parallel equation, is not affected by the Immirzi field. Moreover, in the static weak field limit also the tidal forces does not gain any contribuition depending on $\delta\beta$. Indeed, let us now evaluate the geodesic deviation equation in the comoving frame, \textit{i.e.}:
\begin{equation}
\frac{\partial^2}{\partial \tau^2}\xi^{\alpha}=R\indices{^{\alpha}_{\mu\nu\beta}}U^{\mu}U^{\nu}\xi^{\beta}=R\indices{^{\alpha}_{00\beta}}\xi^{\beta},
\label{geodesic}
\end{equation}
being $\xi=(0,\xi_x,\xi_y,\xi_z)$ a vector denoting the separation between two nearby geodesics and where the entire Riemann tensor is actually given, up to the first order, by:
\begin{equation}
R\indices{^{\rho}_{\mu\sigma\nu}}=\bar{R}\indices{^{\rho}_{\mu\sigma\nu}}+\partial_{\sigma}K\indices{^{\rho}_{\mu\nu}}-\partial_{\nu}K\indices{^{\rho}_{\mu\sigma}}.
\label{RiemK}
\end{equation}
Thus, we claim that, by virtue of \eqref{Kpert} and \eqref{RiemK}, in the Newtonian limit no $\delta\beta$-contribution is present in \eqref{geodesic}, due both the symmetry properties of the contorsion tensor ($\partial_{\beta}K\indices{^{\alpha}_{00}}=0$) and the independence of the Immirzi field from time ($\partial_{0}K\indices{^{\alpha}_{0\beta}}=0$). Therefore, the study of gravitational waves offers the first arena to seek non-trivial effects due the presence of the Immirzi field.
\\ \indent In linearized theory \eqref{beta} reads as
\begin{equation}
\bar{\Box}\delta\beta=0,
\label{betalin}
\end{equation}
\textit{i.e.} the wave equation for a scalar perturbation propagating in vacuum, that can be considered like a genuinely dynamical torsion field. 
\\ \indent Although \eqref{GWonda} and \eqref{betalin} are in principle not coupled, one can easily see as $\delta\beta$ is actually detectable. In fact, let us consider a gravitational plane wave propagating along the $z$ axis in the ordinary transverse and traceless gauge (TT), namely:
\begin{equation}
\tilde{h}_{\mu\nu}=\begin{pmatrix}
0 & 0 & 0 & 0 \\
0 & h^{TT}_{xx} &  h^{TT}_{xy} & 0 \\
0 &  h^{TT}_{xy} & - h^{TT}_{xx} & 0 \\
0 & 0 & 0 & 0 \\
\end{pmatrix}\equiv
\begin{pmatrix}
0 & 0 & 0 & 0 \\
0 & h_{+} &  h_{\times} & 0 \\
0 &  h_{\times} & - h_{+} & 0 \\
0 & 0 & 0 & 0 \\
\end{pmatrix}.
\end{equation}
Since for a TT-wave travelling in z-direction the only non-vanishing components of the Riemann tensor $\bar{R}\indices{^{\rho}_{\mu\sigma\nu}}$ are:
\begin{equation}
\begin{split}
&\bar{R}\indices{^x_{0x0}}=\bar{R}_{x0x0}=-\bar{R}_{y0y0}=-\frac{1}{2}\frac{\partial^2}{\partial t^2}h_{+}\\
&\bar{R}\indices{^y_{0x0}}=\bar{R}_{y0x0}=\bar{R}_{x0y0}=-\frac{1}{2}\frac{\partial^2}{\partial t^2}h_{\times},
\end{split}
\label{hTT}
\end{equation}
inserting \eqref{hTT}, \eqref{Kpert} in \eqref{geodesic} carries out, for a generic $\delta\beta$ wave:
\begin{equation}
\begin{split}
&\frac{\partial^2\xi_{x}}{\partial t^2}=\frac{1}{2}\frac{\partial^2}{\partial t^2}\Big(\xi_x h_{+}+\xi_y h_{\times}\Big)+\frac{1}{2\phi_0}\frac{\partial}{\partial t}\leri{\leri{\vec{\xi}\times\vec{\nabla}}_{x}\delta\beta}\\
&\frac{\partial^2\xi_{y}}{\partial t^2}=\frac{1}{2}\frac{\partial^2}{\partial t^2}\Big(\xi_x h_{\times}-\xi_y h_{+}\Big)+\frac{1}{2\phi_0}\frac{\partial}{\partial t}\leri{\leri{\vec{\xi}\times\vec{\nabla}}_{y}\delta\beta}\\
&\frac{\partial^2\xi_{z}}{\partial t^2}=\frac{1}{2\phi_0}\frac{\partial}{\partial t}\leri{\leri{\vec{\xi}\times\vec{\nabla}}_{z}\delta\beta}.
\end{split}
\label{geodesicimmirzi}
\end{equation}
Then, if we consider a $\delta\beta$ plane wave aligned to the gravitational counterpart, the last one of \eqref{geodesicimmirzi} vanishes identically and we can focus our attention on test masses in the $(x,y)$ plane.
\\ \indent Therefore, be $\delta\beta(t,z)$ given by $(c=1)$:
\begin{equation}
\delta\beta(t,z)=\delta\beta_0\sin(\omega(t-z))
\label{betaz}
\end{equation}
and let us fix $\xi$ as
\begin{equation}
\xi=\leri{0,\xi_x^{(0)}+\delta x,\xi_y^{(0)}+\delta y,0},
\label{xipert}
\end{equation}
being $\xi_x^{(0)},\,\xi_y^{(0)}$ the initial positions and $\delta x,\,\delta y$ the displacements of order $O(h)$ induced by $\delta\beta(t,z)$. Now, if we switch off the truly gravitational modes $h_+,h_{\times}$, a solution for \eqref{geodesicimmirzi} can be easily found:
\begin{equation}
\begin{split}
\delta x(t)&\simeq-\xi_y^{(0)}\frac{\delta\beta_0}{2\phi_0}\sin\omega t\\
\delta y(t)&\simeq\xi_x^{(0)}\frac{\delta\beta_0}{2\phi_0}\sin\omega t,
\end{split}
\label{onlybetasol}
\end{equation}
where we neglected terms of order $O(h^2)$ and we fixed the origin of time such that $\delta\beta=0$ at $t=0$.
\\ \indent The effect of \eqref{onlybetasol} is to rotate a ring of test masses around the $z$-axis without deformation, according an oscillatory behaviour given by the sign of $\sin\omega t$ (Fig.\ref{graficobeta}). Since circular motions are endowed with centripetal forces as well, we expect that the $\beta$-polarization could affect the gravitational modes $h_+,\,h_{\times}$.
Indeed, when $h_+$ is turned off and choosing likewise to  \eqref{betaz} $h_{\times}=h^{(0)}_{\times}\sin\omega t$, \eqref{geodesicimmirzi} shows a solution in terms of effective cross polarizations
\begin{equation}
\begin{split}
\delta x(t)&\simeq\frac{1}{2}\xi_y^{(0)}h^{(-)}_{\times}\sin\omega t\equiv\frac{1}{2}\xi_y^{(0)}h^{(0)}_{\times}\leri{1-\epsilon}\sin\omega t\\
\delta y(t)&\simeq\frac{1}{2}\xi_x^{(0)}h^{(+)}_{\times}\sin\omega t\equiv\frac{1}{2}\xi_y^{(0)}h^{(0)}_{\times}\leri{1+\epsilon}\sin\omega t,
\end{split}
\label{geodesiceff}
\end{equation}
where $\epsilon\equiv\frac{\delta\beta_0}{\phi_0}$.
When $0<\epsilon<1$ holds\footnote{The case $-1<\epsilon<0$ is qualitatively the same up to a counter-clockwise rotation of $\pi / 2$ in the $(x,y)$ plane. Analogous consideration holds for the case $\epsilon\geq 1$.}, the effect of the $\beta$-mode is to change the relative angle between the axes of stretching and shrinking of the $\times$ polarization, that do not turn out to be orthogonal anymore (Fig.\ref{graficoeless1}), but separated by the angle\footnote{Actually $\alpha$ is only one of the angles, being the other one simply given by $\pi-\alpha$.}:
\begin{equation}
\alpha=2\tan^{-1}\leri{\frac{1-\epsilon}{1+\epsilon}}.
\end{equation}
Conversely, if $\epsilon > 1$ the rotational nature of the $\beta$-mode prevails and \eqref{geodesiceff} describe elliptic orbits (Fig.\ref{graficoemore1}), with semi-major axis on $y$, of eccentricity
\begin{equation}
e=\sqrt{\frac{2}{1+\epsilon}}.
\end{equation}
We note as in the limit $\epsilon\to +\infty$ the eccentricity $e$ vanishes, \textit{i.e.} we get circular orbits and the results of \eqref{onlybetasol} are rescued.
\\ \indent Eventually, the value $\epsilon=1$ represents the case of pure shear strains, pointing towards opposite directions according the $x$-half-plan considered (Fig.\ref{graficoe1}).
\\ \indent We conclude this section noting that the results given by \eqref{geodesiceff} are an inherent feature of the first order formulation in the presence of an Immirzi field, regardless the specif nature of the chosen $f(R)$ model. Indeed, it is easy to see how \eqref{geodesiceff} still holds also when $f(R)=R$, simply fixing $\phi_0=1$. Therefore, we stress the fact that the specific signature left by the $\beta$-mode could enable us to distinguish, in principle, a first order formulation from a second order one.
\section{Conclusions}\label{sec5}
In this work, we analyzed the question concerning the first order (Palatini) approach to the $f(R)$ models. 
\\ We addressed the line of thinking that the difficulty to construct a consistent theory when matter couples with the torsion field, suggests the necessity to include \textit{ab initio} torsion in the Lagrangian formulation of the matter gravity system. In particular, we considered a Nieh-Yan term, made no longer a topological contribution by promoting the Immirzi parameter to a real field. 
\\ \indent Our approach allowed to remove torsion from the theory, in terms of the metric and Immirzi field, both in vacuum and in the presence of matter coupled to the torsion field. The obtained theoretical framework resulted into a viable $f(R)$, associated to a new phenomenology of the gravitational field, well traced by the gravitational wave behavior here investigated. 
\\ \indent The possibility to experimentally discriminate between a second order approach to gravity and a first order one, by means of the phenomenology of gravitational wave polarizations, must be regarded as an important perspective for a deeper understanding of the gravity-matter coupling. In fact, it is well known \cite{Wald} that the spinor fields, representing the fermionic matter, are able to generate torsion in a first order formulation, already in the standard Palatini action, but this is not the case if they are treated in a second order metric approach.
\\ \indent The most intriguing phenomenological mark we extracted from the new scenario is the non-orthogonal character of the main axes of the cross-like polarization, for an $f(R)$ theory, as well as for the case $f(R)=R$. 
In this respect, we suggest that it would be a significant analysis to determine the upper
limit of the deviation from $\pi/2$ of the main axes of
the cross polarization, already in gravitational waves
observed events \cite{Abbott:2016blz, Abbott:2016nmj, Abbott:2017vtc}, which could be improved by a systematic
treatment of the stochastic signal.
{\acknowledgments
We would like to thank Fabio Moretti for useful discussions about the effects on the gravitational waves signature due to the Immirzi field propagation.}

\end{document}